# Model-based vs Data-driven Estimation of Vehicle Sideslip Angle and Benefits of Tyre Force Measurements

A. Bertipaglia*, D. de Mol*, M. Alirezaei^, R. Happee* & B. Shyrokau*
*Delft University of Technology, Delft, The Netherlands.
^Eindhoven University of Technology, Eindhoven, The Netherlands.
^Siemens Digital Industries Software, Helmond, The Netherlands.
E-mail: A.Bertipaglia@tudelft.nl

This paper provides a comprehensive comparison of model-based and data-driven approaches and analyses the benefits of using measured tyre forces for vehicle sideslip angle estimation. The model-based approaches are based on an extended Kalman filter and an unscented Kalman filter, in which the measured tyre forces are utilised in the observation model. An adaptive covariance matrix is introduced to minimise the tyre model mismatch during evasive manoeuvres. For data-driven approaches, feed forward and recurrent neural networks are evaluated. Both approaches use the standard inertial measurement unit and the tyre force measurements as inputs. Using the large-scale experimental dataset of 216 manoeuvres, we demonstrate a significant improvement in accuracy using data-driven vs. model-based approaches. Tyre force measurements improve the performance of both model-based and data-driven approaches, especially in the non-linear regime of tyres.

Topics / Identification and Estimation

## 1. INTRODUCTION

The ability to estimate the vehicle sideslip angle in real-time is essential to strengthening the performance of active vehicle control systems [1]. Various driving conditions, such as regular or at the handling limits, steady-state or transient manoeuvres, make the estimation challenging. Significantly, the highly non-linear behaviour of tyres leads to a substantial limitation in the tyre model accuracy. Many solutions for vehicle sideslip angle estimation have been proposed in the past. They can be split into model-based [2], and data-driven [3] approaches. Both allow the incorporation of tyre forces measured by load sensing technology, e.g. load-sensing bearings or smart tyres [4, 5]. This paper provides a comprehensive evaluation of the accuracy of each approach and aims to quantify the benefits of adding tyre force measurements for each of the proposed solutions. Although model-based and data-driven approaches have already been analysed in surveys, their comparison until now is mainly performed on simulation data [6, 7]. Furthermore, we demonstrate the benefits of tyre force measurements for each approach, which is not previously addressed in the literature.

This paper compares the approaches using a large-scale real-world experimental dataset. The dataset contains a great diversity of driving situations. It considers standard vehicle dynamics manoeuvres, e.g. double lane change, slalom, random steer, J-turn, spiral, braking in the turn, and steady-state circular tests, together with recorded laps at the Papenburg track. The dataset includes 216 manoeuvres which correspond to two hours of driving. The log distribution of the sideslip angle and lateral acceleration is represented in Fig. 1. The lower availability of high sideslip angle data points is due to the difficulties of driving in such conditions, and it influences the training of the data-driven approach.

An extended Kalman Filter (EKF) and an unscented Kalman Filter (UKF) using a single-track vehicle model are implemented for the model-based approach. At the same time, a Feed Forward Neural Network (FFNN) and a Recurrent Neural Network (RNN) are considered for the data-driven approach. During the development of the estimators, the longitudinal velocity is assumed to be known, similar to [3, 8].

The contribution of this paper is twofold. First, the accuracy of model-based and data-driven approaches is compared using the large experimental dataset representing various driving manoeuvres. Secondly, the benefit of adding tyre force measurements is assessed for each approach.

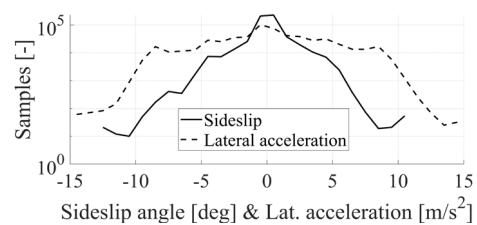

Fig. 1 Log distribution of sideslip angle and lateral acceleration. Each bin corresponds to 1 *deg* or 1 *m/s²*.

## 2. MODEL-BASED APPROACH

The model-based approach uses the physical knowledge of a vehicle model for state estimation. Open-loop deterministic models are insufficient to provide an accurate estimation due to i) the mismatches between the physical and modelled vehicle behaviour, ii) the uncontrolled disturbances and the stochastic measurement noise [9]. Thus, stochastic closed-loop observers, e.g. EKF, UKF, and particle filters, are currently applied to estimate unknown variables. EKF and UKF are the current industrial state-of-art for vehicle sideslip angle estimation because the accuracy of these approaches can be guaranteed in a specific operating region [10], and their properties can be easily assessed [11]. Both algorithms assume the process and observation noise parameters as uncorrelated and Gaussian.

### 2.1 Extended Kalman Filter

The EKF is the most widely used non-linear observer based on the Kalman filter. It linearises about the estimate of the current mean and covariance of the non-linear stochastic model to compute the propagated covariance and mean. The EKF linearises up to the 1$^{st}$ term of the Taylor series expansion and corresponds to the Jacobian matrix calculation:

$$A_t = \frac{\partial f\left(\overline{X}_{t-1|t-1}, u_t, 0\right)}{\partial X} \quad (1)$$

where $f(...)$ is the stochastic model, $X$ are the states, $\overline{X}$ represents the estimated ones, $u$ is the input, and $t$ is the time. Thus, it requires an analytical derivation of the Jacobian matrices, which can be complex and time-consuming. Furthermore, when the process model is highly non-linear and subjected to strong uncertainties, EKF can become unstable. This can also happen when EKF cannot capture the non-linearities due to local linearisation. This problem is visually represented in Fig. 2. The propagated covariance cannot capture the non-linearities of the tyre force subjected to high uncertainty of the tyre slip angle.

### 2.2 Unscented Kalman Filter

The UKF is based on the unscented transformation, which assumes easier an approximation of a probability distribution rather than an arbitrary non-linear function [12]. Thus, the propagated covariance and mean are computed from a sigma point ($\sigma^{(s)}$) cloud propagated through the non-linear process model. After the non-linear function has been applied to each of the $\sigma^{(s)}$, it is possible to approximate the propagated mean ($\overline{X}$) and covariance ($P$) as follows:

$$\overline{X}_{t+1|t} = \sum_{i=0}^{2n} \omega_{i,t|t-1}^{(m)} \sigma_{i,t|t-1}^{(s)} \quad (2)$$

$$P_{t+1|t} = \sum_{i=0}^{2n} \omega_i^{(c)} \left[\sigma_{i,t+1|t}^{(s)} - \overline{X}_{t+1|t}\right]\left[\sigma_{i,t+1|t}^{(s)} - \overline{X}_{t+1|t}\right]^T \quad (3)$$

where $n$ is the number of $\sigma^{(s)}$, $\omega^{(m)}$ and $\omega^{(c)}$ are the sigma point weights for the mean and the covariance, respectively. The $\sigma^{(s)}$ and their weights are deterministically chosen to enhance the unscented transformation performance. Thus, a linearisation up to a 3$^{rd}$ order term of the Taylor series expansion can be achieved. If the model is strongly non-linear and subjected to high uncertainties, the broader scattering of $\sigma^{(s)}$ will allow to better capture the non-linearities, see Fig. 2. The UKF does not need an analytical derivation of the Jacobian matrices, thanks to the unscented transformation.

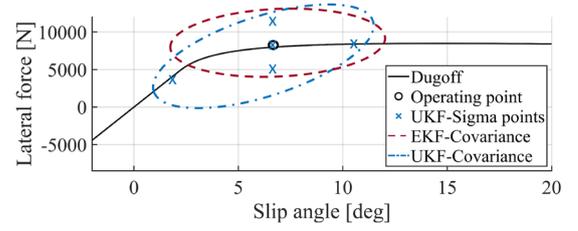

Fig. 2 Propagated covariance by EKF and UKF.

### 2.3 Vehicle Model

The single-track model with tyre axle forces computed by the Dugoff tyre is chosen in this study. The vehicle states are the lateral velocity ($V_y$) and the yaw rate ($\dot{\psi}$), while the inputs are the road wheel angle ($\delta$) and the longitudinal velocity ($V_x$). Two different vehicle measurement sets are considered: the first ($Y_1$) is formed by lateral acceleration ($a_y$) and the $\dot{\psi}$, and the second ($Y_2$) is formed by the same measurements of $Y_1$ plus the lateral tyre forces, front ($F_{yF}$) and rear ($F_{yR}$). The stochastic process model ($\dot{X} = f(X,u) + \omega$) is represented as follows:

$$f(X,u) = \begin{cases} \dot{V}_y = \frac{1}{m}\left(F_{yF}(X,u)\cos(\delta) + F_{yR}(X,u)\right) - V_x\dot{\psi} \\ \ddot{\psi} = \frac{1}{I_{zz}}\left(L_F F_{yF}(X,u)\cos(\delta) - L_R F_{yR}(X,u)\right) \end{cases} \quad (4)$$

where $m$ stands for the vehicle mass (1970 kg), $I_{zz}$ is the vehicle moment of inertia around the vertical axis (3498 kg m2), LF and LR are the geometrical distances between the centre of gravity and the front (1.47 m) and rear (1.41 m) axle respectively. $\omega$ is the vector containing the process noise parameters $\sigma_{V_y}$ and $\sigma_{\dot{\psi}}$, which compensate for the model mismatch to the actual vehicle and the discretisation error. The filter performance is strongly connected with these parameters, so they are tuned using a two-stage Bayesian optimisation (TSBO) [13].

The observation model ($Y = g(X,u) + v$) aims to compare the process model predictions with the available measurements. It is defined as follows:

$$g(X,u) = \begin{cases} a_{y,me} = \frac{1}{m}\left(F_{yF}(X,u)\cos(\delta) + F_{yR}(X,u)\right) \\ \dot{\psi}_{me} = \dot{\psi} \\ F_{yF,me} = F_{yF}(X,u) \\ F_{yR,me} = F_{yR}(X,u) \end{cases} \quad (5)$$

Where $v$ is the vector containing the observation noise parameters $\sigma_{a_y,me}$, $\sigma_{\dot{\psi},me}$, $\sigma_{F_{yF}}$ and $\sigma_{F_{yR}}$ of respectively the $a_y$, $\dot{\psi}$, $F_{yF}$ and $F_{yR}$ measurements. Eq. 5 is the observation model for $Y_2$; the one for $Y_1$ would be formed only by the first two equations of the system. The observation noise parameters are tuned by statistical analysis of the vehicle sensor measurements. An adaptive noise parameter technique is applied to $\sigma_{F_{yF}}$ and $\sigma_{F_{yR}}$ to enhance the filter performance. The adaptivity is triggered when the vehicle behaves non-linearly to give more trust to the tyre force measurements [8]. This is measured through the absolute difference between the measured tyre forces and the predicted ones ($\Delta F_y$). When it is higher than a selected threshold ($Tr = 700$ N), the adaption is triggered and to avoid the chattering phenomenon, a hysteresis loop is implemented (Fig. 3, left). The adaptivity consists of reducing the level of noise associated with tyre force measurements to increase the value of the feedback Kalman gain. Thus, state estimation is sped up during manoeuvres at the handling limit and does not follow noise during normal driving situations. $\sigma_{F_yF}$ and $\sigma_{F_yR}$ are reduced according to the following equation:

$$\sigma_{adap} = \sigma_{nom} - r_{\max}\left(1 - e^{-0.05\left(\frac{\Delta F_y - Tr}{\sigma}\right)^2}\right) \quad (6)$$

where $\sigma_{nom}$ is the nominal value of $\sigma_{F_yF}$ or $\sigma_{F_yR}$, $r_{max}$ (12.5 N) is the maximum reduction possible, and $\sigma$ is a parameter which defines the slope steepness (800). A visual representation of Eq. 6 is shown in Fig. 3 right. All the user-defined parameters are tuned using a TSBO.

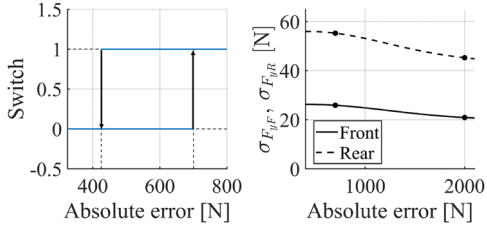

Fig. 3 Left: hysteresis loop for $\sigma_{F_yF}$ and $\sigma_{F_yR}$. Right: shape of the adaptive $\sigma_{F_yF}$ and $\sigma_{F_yR}$.

## 3. DATA-DRIVEN APPROACH

A model-based approach is restricted by model assumptions and model parametrisation. The data-driven approach is model-free and relies only on the universal approximator property of neural networks (NN). However, a data-driven approach requires high-quality data points to perform correctly. Furthermore, its properties are almost impossible to assess [11]. This work analyses two of the most common NN architectures: FFNN and RNN. Both are compared using two input sets. The first one ($I_1$) considers only measurements from the inertial measurement unit (IMU), which are $V_x$, longitudinal acceleration ($a_x$), $a_y$, $\dot{\psi}$ and $\delta$. The second input set ($I_2$) considers the same five measurements of $I_1$ plus the longitudinal, lateral and vertical tyre forces for each of the four wheels. Thus, in total, seventeen measurements are used. In the pre-processing phase, the input measurements are normalised because each input has a different physical meaning and order of magnitude. All the inputs are mapped onto the interval [0, 1] to speed up and stabilise the training process.

### 3.1 Feed Forward Neural Network

FFNN is the most straightforward NN regarding implementation. It usually consists of an input layer, a couple of hidden layers and an output layer. The input only moves forward, so the time correlation information does not have any effect. The FFNN based on $I_1$ is formed by two hidden layers with 250 and 100 neurons each and Rectified Linear Unit (ReLU) activation functions, while the FFNN based on $I_2$ is formed by two hidden layers with 150 and 50 neurons each. Both FFNNs use a dropout regularisation technique equal to 0.2 and a Xavier weight initialisation to avoid overfitting.

Furthermore, an early stopping method with patience equal to 20 is applied to avoid overfitting. A mean squared error (MSE) loss function is used for the training, and its gradients for NN parameters are computed following the back-propagation algorithm. The loss function is minimised using a mini-batch stochastic gradient descent algorithm based on a standard ADAM optimiser with a learning rate of 0.0008. The mini-batch size is 1024. The training procedures' user-defined parameters are optimised through a Bayesian optimisation.

### 3.2 Recurrent Neural Network

Measurements at previous time steps possess predictive power on the current sideslip angle, so a RNN architecture is considered. A long short-term memory (LSTM) cell is used to avoid the vanishing/exploding gradient problem. The RNN based on $I_1$ uses two hidden layers of 100 and 80 LSTM cells. The RNN based on $I_2$ uses two hidden layers of 100 and 50 LSTM cells. The first layers use a hyperbolic tangent activation function, while the second uses a sigmoid activation function. The LSTM time window is 0.20 s. Both RNNs use a dropout regularisation technique equal to 0.2 and a Xavier weight initialisation to avoid overfitting. An early stopping method with patience equal to 4 is applied to avoid overfitting. A MSE loss function is applied for the training, and its gradients for NN parameters are computed following the back-propagation through time algorithm. The loss function is minimised using a mini-batch stochastic gradient descent algorithm based on a NADAM optimiser with a learning rate of 5e-4. The mini-batch size is 256. All the user-defined parameters of the training procedures are optimised through Bayesian optimisation.

## 4. EXPERIMENTS
### 4.1 Dataset

This work uses an experimental dataset recorded at the Automotive Testing Papenburg GmbH with a BMW Series 3. The vehicle was instrumented with the standard IMU, wheel force transducers and load sensing bearings for each wheel, GPS and a Corrsys-Datron optical sensor to measure the sideslip angle (accuracy $\pm 0.2$ deg). The optical speed sensor measurement is used as ground truth to train the NN and tune the EKF/UKF. The tyre forces used in the dataset are from the wheel force transducers because they are more common sensors in the research. However, load-sensing bearings demonstrate a similar accuracy.

The dataset consists of 216 manoeuvres covering standard vehicle dynamics manoeuvres, e.g. double lane change, slalom, random steer, J-turn, spiral, braking in the turn, and steady-state circular tests, together with recorded laps at the Papenburg track. The vehicle was driven on dry asphalt with tyres inflated according to the manufacturer's specifications. Two different settings (On, Off) of electronic stability control were used.

All signals were recorded at 100 Hz, the standard frequency for vehicle state estimation. The measurements are considered only when the $V_x$ is higher than 2.5 m/s. A statistical outlier removal has been applied to remove extreme outliers. However, particular attention is paid not to delete edge case measurements which are the most precious data. Regardless, all the manoeuvres are visually inspected.

For the data-driven approach, the measurements are filtered using a low-pass zero-phase filter with a cut-off frequency of 5 $Hz$ [14]. The filter is designed using a finite impulse response technique.

The dataset is split into three subsets: training (75%), validation (15%) and test (10%). The test set contains manoeuvres representing the entire driving behaviour, but more focus is paid to highly non-linear situations. It includes 23 manoeuvres: 2 braking in the turn, 2 skidpad, 5 J-turn, 4 slalom, 4 lane change, 2 random steers, 1 lap track and 3 spiral. The NN training and the EKF/UKF tuning use only the data from the training and the validation set.

**4.2 Key Performance Indicators**

The performance of the different approaches for vehicle sideslip angle estimation is assessed through four key performance indicators (KPIs).

- The root mean squared error (RMSE) assesses the overall estimation performance as follows:

$$RMSE = \sqrt{\sum_{i=1}^{L} \frac{\left(\bar{\beta}_i - \beta_i\right)^2}{L}} \quad (7)$$

where $L$ is the length of a manoeuvre, and $\beta$ is the sideslip angle.

- The non-linear RMSE ($RMSE_{nl}$) corresponds to the RMSE computed only when the $|a_y| \geq 4$ $m/s^2$. It measures the estimation performance when the vehicle behaves non-linearly.
- The absolute maximum error (ME) measures the worst estimation performance.
- The non-linear ME ($ME_{nl}$) measures the worst estimation performance when the vehicle behaves non-linearly.

**4.3 Results**

The hereafter analysis is conducted on the test set. The overall comparison of the model-based approaches is presented in Table 1. The comparison between EKF and UKF shows that the latter outperforms the EKF for all four KPIs. The UKF improvement of the $RMSE_{nl}$ (12.1%) is particularly relevant. It highlights how the UKF behaves better when the vehicle is in a non-linear operating region. This is not only visible from the $RMSE_{nl}$, but also from the improvement in the $ME_{nl}$ (21.8%). The UKF's improved performance finds an explanation in the aforementioned linearisation technique.

Interestingly, the EKF with tyre force measurements reaches an overall performance very close to the UKF. The reason is that the tyre force measurements help reduce model mismatches, especially when the vehicle behaves non-linearly. Fig. 4 shows the experimental and estimated $F_{yF}$. It is visible that when the $a_y$ is higher than 5 $m/s^2$ (indicated by the dashed threshold), the tyre model mismatch is growing. However, the EKF with tyre force measurements has a lower mismatch than a EKF with only IMU measurements. Moreover, the difference between estimated and measured tyre force triggers the observation noise parameters adaptation. The latter pushes the Kalman gain's magnitude and improves the sideslip angle estimation.

On average, the UKF with tyre force measurements outperforms all other model-based algorithms for all four KPIs. Despite this, the tyre force measurements benefits for the UKF are lower than for the EKF because the first has a better ability to deal with vehicle non-linearities.

A comparison between the four model-based algorithms is presented in Fig. 5. It shows the sideslip angle estimation in a J-turn manoeuvre at the handling limit. The UKF with tyre force measurements better estimates the sideslip angle, especially in the peak area (from 3.5 s to 4.5 s). The EKF with tyre force measurements performs similarly to the corresponding UKF but cannot estimate the peak correctly.

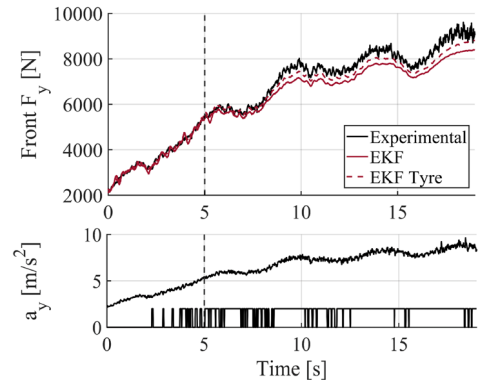

Fig. 4 Skidpad manoeuvre. Top: measured and estimated front tyre lateral force. Bottom: vehicle lateral acceleration and observation noise parameters flag.

Table 1 Model-based approach comparison.

| KPIs [$deg$] | EKF | EKF Tyre | UKF | UKF Tyre |
|---|---|---|---|---|
| RMSE | 0.421 | 0.391 | 0.394 | **0.370** |
| $RMSE_{nl}$ | 0.563 | 0.488 | 0.490 | **0.448** |
| ME | 1.368 | 1.257 | 1.180 | **1.113** |
| $ME_{nl}$ | 1.271 | 1.169 | 1.068 | **0.994** |

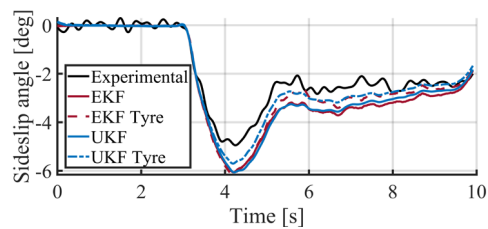

Fig. 5 J-turn manoeuver. Comparison of the sideslip angle estimation between the model-based approaches.

The overall comparison of the data-driven approaches is summarised in Table 2. FFNN and RNN without tyre force measurements have similar performances, but some crucial outcomes are noticeable. The overall RMSE of the FNNN is lower (3.4%) than the RNN one. This result is counterintuitive because the time information exploited by the RNN contains some predictive power. However, when the RMSE$_{nl}$ of the RNN is compared with the one of the FFNN, it results in an improvement of a 13.2%. Thus, the RNN predictive power is exploited when the vehicle behaves non-linearly. On the contrary, in a linear operating region, the simplest structure of a FFNN results beneficial compared to the highest number of parameters of a RNN with LSTM cells. Similar conclusions are visible for the ME and ME$_{nl}$ analysis.

If the tyre force measurements are included in the NN input sets, the overall RMSE performance of the FFNN and RNN improves by, respectively, 44.9% and 42.6%. This is coherent with the literature because tyre force measurements contain precious information to describe vehicle dynamics. The benefits of tyre force measurements are even more visible when the RMSE$_{nl}$ is compared. The overall RMSE performance of the FFNN and RNN improves by 68.1% and 58.2%, respectively. The reason is that the data-driven approach with the standard IMU measurements was poorly performing when the vehicle was in a non-linear operating region due to a lower amount of data in the training set. If the number of recorded manoeuvres is not increased with the tyre force measurements, vice versa, the amount of information is highly increased. Particularly important is the fact that a simple FFNN reaches a better performance than a RNN when the tyre force measurements are included in the input set. This is explained by the fact that the RNN prediction power is insufficient to compensate for the higher numbers of parameters to be trained.

Table 2 Data-driven approach comparison.

| KPIs [deg] | FFNN | FFNN Tyre | RNN | RNN Tyre |
| --- | --- | --- | --- | --- |
| RMSE | 0.379 | **0.209** | 0.392 | 0.225 |
| RMSE$_{nl}$ | 0.645 | **0.206** | 0.560 | 0.234 |
| ME | 1.635 | 0.784 | 1.649 | **0.783** |
| ME$_{nl}$ | 1.509 | **0.592** | 1.495 | 0.643 |

The log distribution of the sideslip angle error for all the data-driven NNs is represented in Fig. 6. It highlights how the FFNN and RNN with tyre force measurements have a smaller standard deviation than the respective NN with only IMU measurements. Furthermore, it highlights how FFNN and RNN are subjected to high ME and ME$_{nl}$. A possible explanation is that NN learns only from the data, so the estimations are not validated by a vehicle model.

A comparison between the data-driven approaches is presented in Fig. 7. It shows the sideslip angle estimation in a J-turn manoeuvre at the handling limit. The FFNN and the RNN with tyre force measurements better estimate the sideslip angle, especially in the peak area (from 2.5 s to 3.5 s). Moreover, the FFNN with tyre force measurements has a perfect estimation of the following J-turn section (from 4 s to 10 s), while the corresponding RNN has a slight bias. Both NNs with only IMU measurements cannot capture the sideslip angle peak due to the limited data in the training set, representing this condition.

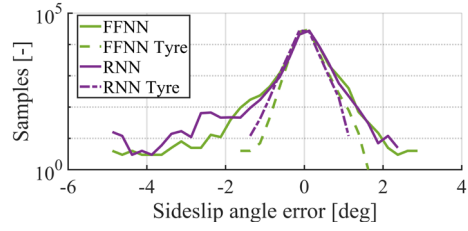

Fig. 6 Distribution of the sideslip angle error for every data-driven approach. Each bin is 0.25 *deg* wide.

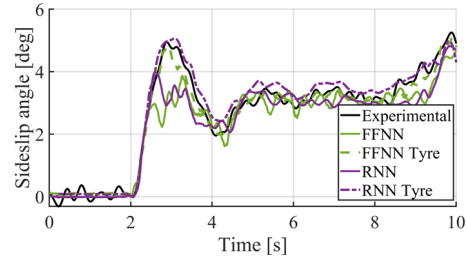

Fig. 7 J-turn manoeuver. Comparison of the sideslip angle estimation between the data-driven approaches.

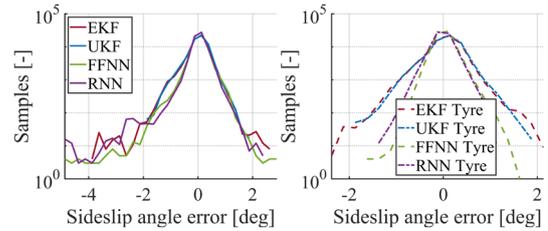

Fig. 8 Left: log distribution of the sideslip angle error for every approach with IMU measurements. Right: log distribution of the sideslip angle error for every approach with tyre force measurements. Each bin is 0.25 *deg* wide.

Below the model-based and the data-driven approaches are compared using Table 1 and Table 2. At first, both approaches are compared when only IMU measurements are available. The RMSE of the data-driven approaches are lower than the model-based approaches, but the RMSE$_{nl}$ shows an opposite trend. The reason is that data-driven approaches are dependent on the quality/quantity of the data. They are less when the vehicle behaves non-linearly, and the sideslip angle is high (>*5 deg*), see Fig. 1, so the performance of the data-driven approach decreases in this situation. Moreover, data-driven approaches are more prone to high ME and ME$_{nl}$ than model-based approaches. This can be also seen in the log distribution of the sideslip angle error, (Fig. 8, left). EKF, FFNN and RNN present few very high sideslip angle errors (>3 *deg*), while UKF has a shorter range. However, the data-driven approach has more samples in the small error bins (>1 *deg*).

A different conclusion emerges if both approaches are compared when adding tyre force measurements. In this case, the data-driven approach outperforms the model-based one for all four KPIs. Thus, the tyre force measurements majorly improve the data-driven approach's performance. This can also be observed in the log distribution of the sideslip angle error (Fig. 8, right). The standard deviation of the FFNN and RNN is clearly lower than the one of the model-based approach.

Fig. 9 shows the sideslip angle estimation in a J-turn manoeuvre at the handling limit with tyre force measurements. The overall performance of the considered algorithms is similar, but the FFNN has a more accurate estimation. Similar considerations can be seen in Fig. 10. On the right, a slalom manoeuvre is represented, while on the left, a portion of the Papenburg track. Fig. 10 shows how both approaches reach a very high estimation accuracy also in a very high dynamic manoeuvre and when the sideslip angle peaks up to 11 *deg*.

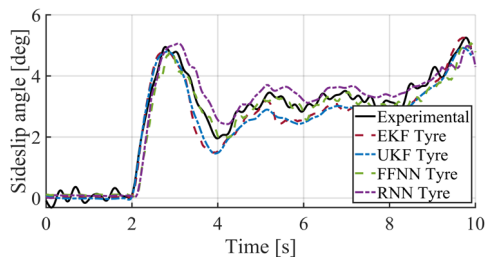

Fig. 9 J-turn manoeuver. Sideslip angle estimation comparison between approaches with tyre force measurements.

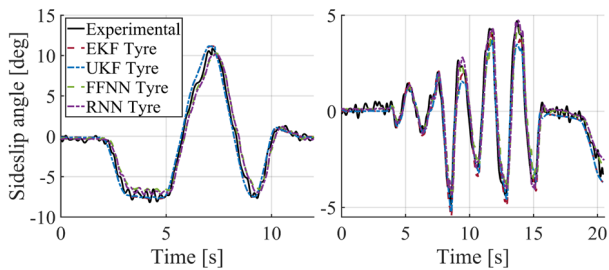

Fig. 10 Left: Papenburg track corners. Right: slalom manoeuver. Comparison of the estimated sideslip angle between the approaches with tyre force measurements.

## 5. CONCLUSIONS

This paper presented a comprehensive comparison between model-based and data-driven approaches for vehicle sideslip angle estimation. Moreover, the benefits of adding measured tyre forces are demonstrated. Using an extensive experimental dataset, it has been shown that the UKF is more accurate than the EKF, but their performance becomes similar with the usage of additional tyre force measurements. FFNN and RNN have similar performance, and the RNN prediction power is perceptible only when the vehicle behaves non-linearly. When the data-driven approaches rely on the tyre force measurements, it strongly outperforms the model-based approach. In this situation, the simpler structure of the FFNN yields a better estimation than the RNN. However, the data-driven approach can still suffer from the low amount of data in the training set. Future work involves combining the pros of the model-based and data-driven approach to develop a hybrid approach for vehicle sideslip angle estimation.

## ACKNOWLEDGEMENT

The Dutch Science Foundation NWO-TTW supports the research within the EVOLVE project (nr. 18484).